\documentclass[
aps,%
11pt,%
final,%
notitlepage,%
oneside,%
twocolumn,%
nobibnotes,%
nofootinbib,%
superscriptaddress,%
noshowpacs,%
centertags]%
{revtex4}

\usepackage{multirow}

\begin{document}

\selectlanguage{english}
       
       
\title{KDG\,218, a Nearby Ultra-Diffuse Galaxy}


\author{\firstname{I.~D.}~\surname{Karachentsev}}
\author{\firstname{L.~N.}~\surname{Makarova}}
\author{\firstname{M.~E.}~\surname{Sharina}} \affiliation{\saoname}

\author{\firstname{V.~E.}~\surname{Karachentseva}}
\affiliation{Main Astronomical Observatory, National Academy of Sciences of Ukraine, 03143 Kyiv, Ukraine}

\received{September 26, 2017 } \revised{October 2, 2017 }

\onecolumngrid
{\scriptsize
Original Russian Text @ I.D.~Karachentsev,
L.N.~Makarova, M.E.~Sharina, V.E.~Karachentseva 2017,\\
published in Astrofizicheskii Byulleten, 2017, Vol.72, No.4, pp.413-421}

\begin{abstract}
We present properties of the low-surface-brightness galaxy KDG\,218
observed with the HST/ACS. The galaxy has a half-light (effective)
diameter of $a_e =47$\arcsec\ and a central surface brightness of
$SB_V(0) = 24\fm4/\sq\arcsec$. The galaxy remains unresolved with the
HST/ACS, which implies its distance of \mbox{$D > 13.1$~Mpc} and
linear effective diameter of $A_e >3.0$~kpc. We notice that KDG\,218
is most likely associated with a galaxy group around the massive
lenticular NGC\,4958 galaxy at approximately $22$~Mpc, or with the
Virgo Southern Extension filament at approximately $16.5$~Mpc. At
these distances, the galaxy is classified as an ultra-diffuse galaxy
(UDG) similar to those found in the Virgo, Fornax, and Coma clusters.
We also present a sample of 15 UDG candidates in the Local Volume.
These sample galaxies have the following mean parameters: $\langle
D\rangle = 5.1$~Mpc, $\langle A_e\rangle = 4.8$~kpc, and $\langle
SB_B (e) \rangle = 27\fm4/\sq\arcsec$. All the local UDG candidates
reside near massive galaxies located in the regions with the mean
stellar mass density (within 1~Mpc) about 50 times greater than the
average cosmic density. The local fraction of UDGs does not exceed
1.5\% of the Local Volume population. We notice that the presented
sample of local UDGs is a heterogeneous one containing irregular,
transition, and tidal types, as well as objects consisting of an old
stellar population.

\end{abstract}

\maketitle

\section{INTRODUCTION}
Basing on reproductions of the first Palomar Sky Survey (POSS-I),
van~den~Bergh searched for low-surface-brightness dwarf galaxies. The
summary list of these objects~\cite{ber1959:Karachentsev_n_en}, called DDO galaxies,
contains 222 dwarf systems mainly late-types: Ir, Im, Sm with a
median radial velocity of about $1200$~km\,s$ ^{-1}$. Later on,
Karachentseva~\cite{kara1968:Karachentsev_n_en} carried out a search for fainter dwarf
galaxies in POSS-I; the number of these objects (KDG) was 241. The
typical surface brightness of KDG objects turned out
\mbox{1--2}~magnitudes fainter than that for the DDOs, and the
majority of them were classified as smooth spheroidal systems (dSph)
without a young population. The summary catalog of
low-surface-brightness dwarf galaxies extended with the results of
new to date all-sky surveys contained about
1500~objects~\cite{kara1988:Karachentsev_n_en}. The sky distribution shows that most
galaxies from the catalog are the members of the Local Supercluster
and strongly concentrated towards the Virgo cluster.

Substitution of the photographic plates on CCD detectors resulted in
detection of an enormous number of new extremely
low-surface-brightness dwarf galaxies in the range of
$SB\simeq[25$--$27]^{\rm m}/\sq\arcsec$. The typical average surface
brightness in the nearest dwarf systems resolved into stars reaches
approximately $[28$--$30]^{\rm m}/\sq\arcsec$.

Generally, normal and dwarf galaxies in the wide range of the
absolute magnitudes $M$ follow the relation $SB \sim M/3$ which
reflects the approximate consistency of the mean spatial luminosity
in massive and dwarf galaxies~\cite{kar2013:Karachentsev_n_en}. However, recently a new
population of extremely low-surface-brightness galaxies has been
found, the luminosity of which is typical of dwarf systems
\mbox{($M_B>-15^{\rm m}$)}, and sizes are comparable with those of
normal galaxies. This galaxy category was called {\it ultra-diffuse
galaxies} (UDG). According to the paper~\cite{dok2015a:Karachentsev_n_en}, galaxies
with the effective diameter $A_e>3.0$~kpc and the central surface
brightness in $g$ band $SB_g(0)>24^{\rm m}/\sq\arcsec$ should be
referred to UDGs. The same authors~\cite{dok2015b:Karachentsev_n_en} detected about 50
UDG candidates in the rich Coma cluster and confirmed with
spectroscopic observations that they really belong to the cluster. In
the last two years, ultra-diffuse galaxies were found in the nearby
Virgo cluster~\cite{mih2015:Karachentsev_n_en,mih2017:Karachentsev_n_en}, in the Fornax
cluster~\cite{mun2015:Karachentsev_n_en}, in the Perseus
supercluster~\cite{mar2016:Karachentsev_n_en,wit2017:Karachentsev_n_en}, and in some other
clusters~\cite{bur2016:Karachentsev_n_en}. The authors of the paper~\cite{kod2015:Karachentsev_n_en}
reported on the detection of at least 800 UDG candidates in the
central region of the Coma cluster. The presence of ultra-diffuse
galaxies in the local groups around Cen~A, NGC~253, and NGC~5485 was
also shown in~\mbox{\cite{crn2016:Karachentsev_n_en,tol2016:Karachentsev_n_en,mer2016:Karachentsev_n_en}.} According to
paper~\cite{rom2017:Karachentsev_n_en}, about 40\% of UDGs belong to clusters, about
20\%---to groups, and about 40\%---to scattered filaments, while in
the general field such objects are almost not found.

\begin{figure*}[tbp!] \setcaptionmargin{5mm} \onelinecaptionsfalse \captionstyle{normal}

\includegraphics[scale=0.65]{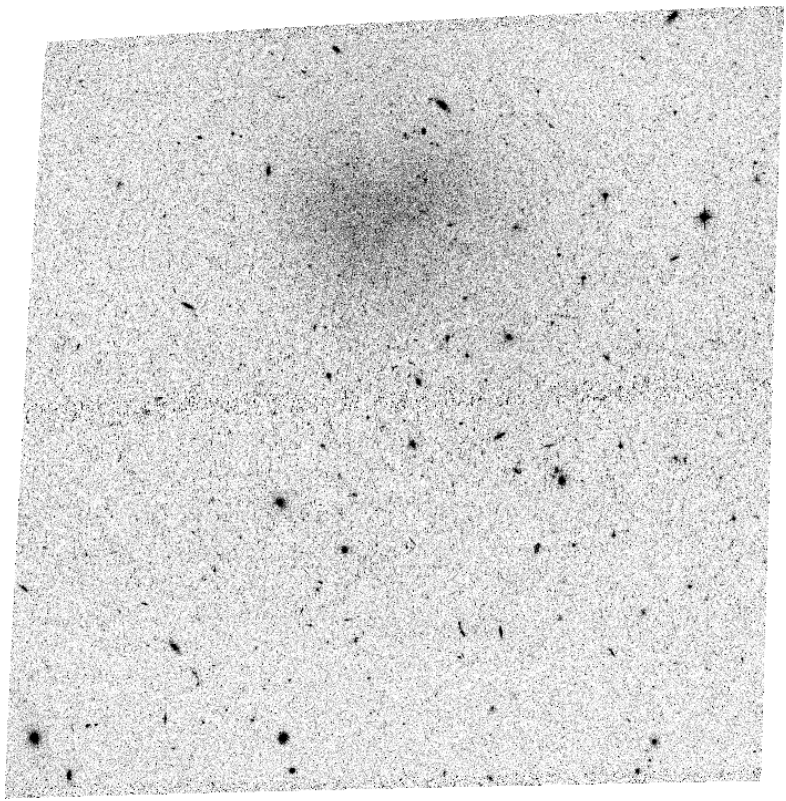}
\caption{Image of the KDG\,218 galaxy obtained with the HST/ACS in
the F606W filter. The field size is about $3\farcm5\times3\farcm5 $,
the North is on the right, the East is at the top.} \end{figure*}

In this paper, we present arguments of the fact that the
low-surface-brightness galaxy KDG\,218 located on the outskirts of
the Virgo cluster may be assigned  to UDG-type systems.

\section{OBSERVED PROPERTIES OF KDG\,218}

\subsection{Observations with the HST/ACS}

\begin{figure*} \setcaptionmargin{5mm} \onelinecaptionstrue \captionstyle{normal}
\includegraphics[scale=0.54]{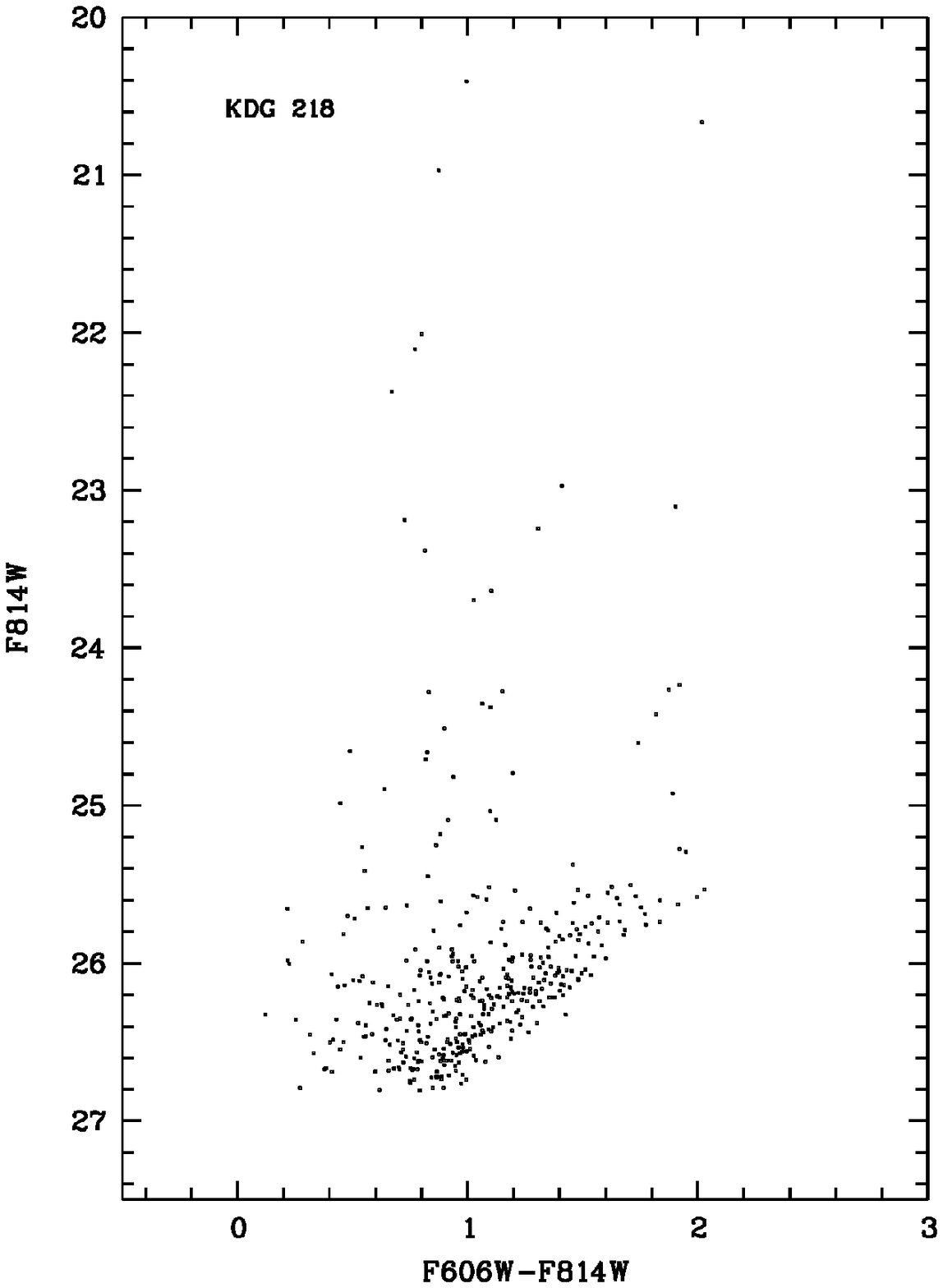}
\hspace{5mm}
\includegraphics[scale=0.54]{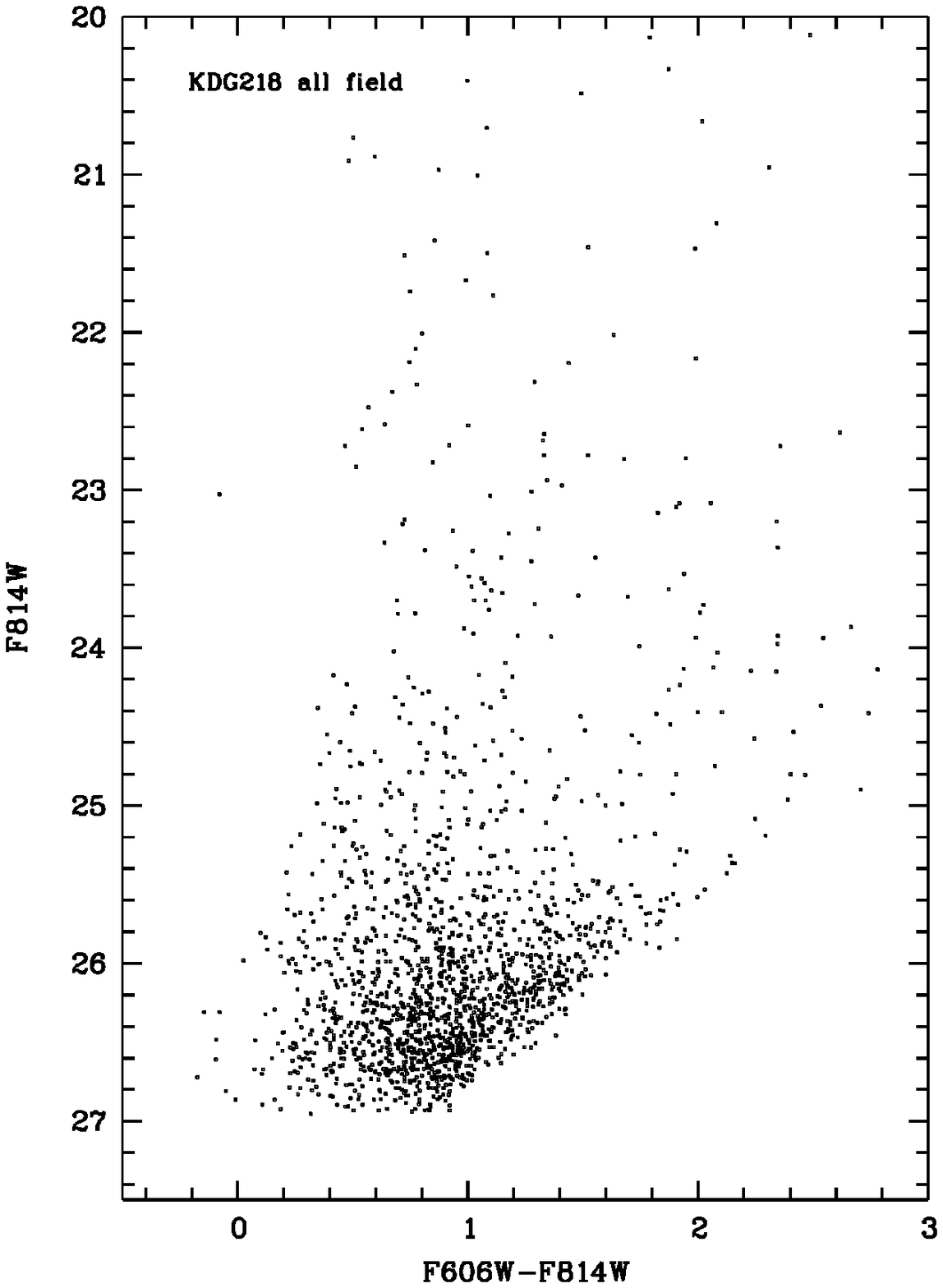}
\caption{Color--magnitude diagrams in the central region of KDG\,218
of a$1\arcmin\times 1\arcmin$ size (left) and in the whole ACS image
(right).} \end{figure*}

The low-surface-brightness galaxy KDG\,218 with the equatorial
coordinates \mbox{$13^{\rm h}05^{\rm
m}43\fs9$\,--\,$07\degr45\arcmin32\arcsec$} (J2000.0) was included in
the program of distance measurements for galaxies located on the
front border of the Virgo cluster (the program GO-14636, PI
I.~D.~Karachentsev). The KDG\,218 images were obtained on May 18,
2017 using the Advanced Camera for Survey (ACS) at the Hubble Space
Telescope (HST) in the F606W (broad-band $V$) and F814W (broad-band
$I$) filters with integrated exposures of 1030~s in each filter.
Figure~1 shows the image of the galaxy in the F606W filter. In the
Updated Nearby Galaxy Catalog (=UNGC~\cite{kar2013:Karachentsev_n_en}), KDG\,218 is
classified as a dwarf system of the transition type Tr with the
integral magnitude $B=16\fm8$, angular diameter $1\farcm8$, and
average surface brightness \mbox{$SB=26\fm6/\sq\arcsec$}. We carried
out stellar photometry using the DOLPHOT~\cite{dol2002:Karachentsev_n_en} package.
Figure~2 presents color-magnitude diagrams (CMD) for the detected
stars within the galaxy and in the whole ACS field.  As we can see, KDG\,218 is almost unresolved into
stars. Inside the galaxy, there are no blue stars brighter than
$I\simeq25^{\rm m}$, and redder stars can be considered as foreground
and background objects. We do not find the presence of RGB stars in
the CMD. Some concentration of stars near the photometric limit in
the F814W and F606W filters is caused by the photometric errors which
is confirmed with simulation of artificial stars with DOLPHOT.

Assuming that the red giant branch tip with $M_I(\rm TRGB)=-4\fm05$
and $V-I\simeq+1.0$ is not higher than $I\simeq 26\fm6$, we obtain
the distance module estimate for KDG\,218 $(m-M)_0>30.58$. According
to the paper~\cite{sch2011:Karachentsev_n_en}, \mbox{$A_I=0\fm07$} is accepted for the
Galactic extinction here. Thus, the KDG\,218 galaxy is at a distance
not closer than $13.1$~Mpc.

Let us notice that the distribution of star-like objects selected
using DOLPHOT with apparent magnitudes of \mbox{$26.5>I>26.0$} and
color indices of \mbox{$1.7>V-I>0.8$} shows weak concentration
towards the center of KDG\,218. Abundance of such objects inside the
galaxy is \mbox{$\Delta N=62\pm11$.} Possibly, some part of this
abundance is determined by the presence of globular clusters in
KDG\,218, luminosity and color of which correspond to the selected
intervals.

\subsection{Surface Photometry}

\begin{figure*} \setcaptionmargin{5mm} \onelinecaptionstrue \captionstyle{normal}
\includegraphics[scale=0.6,angle=-90]{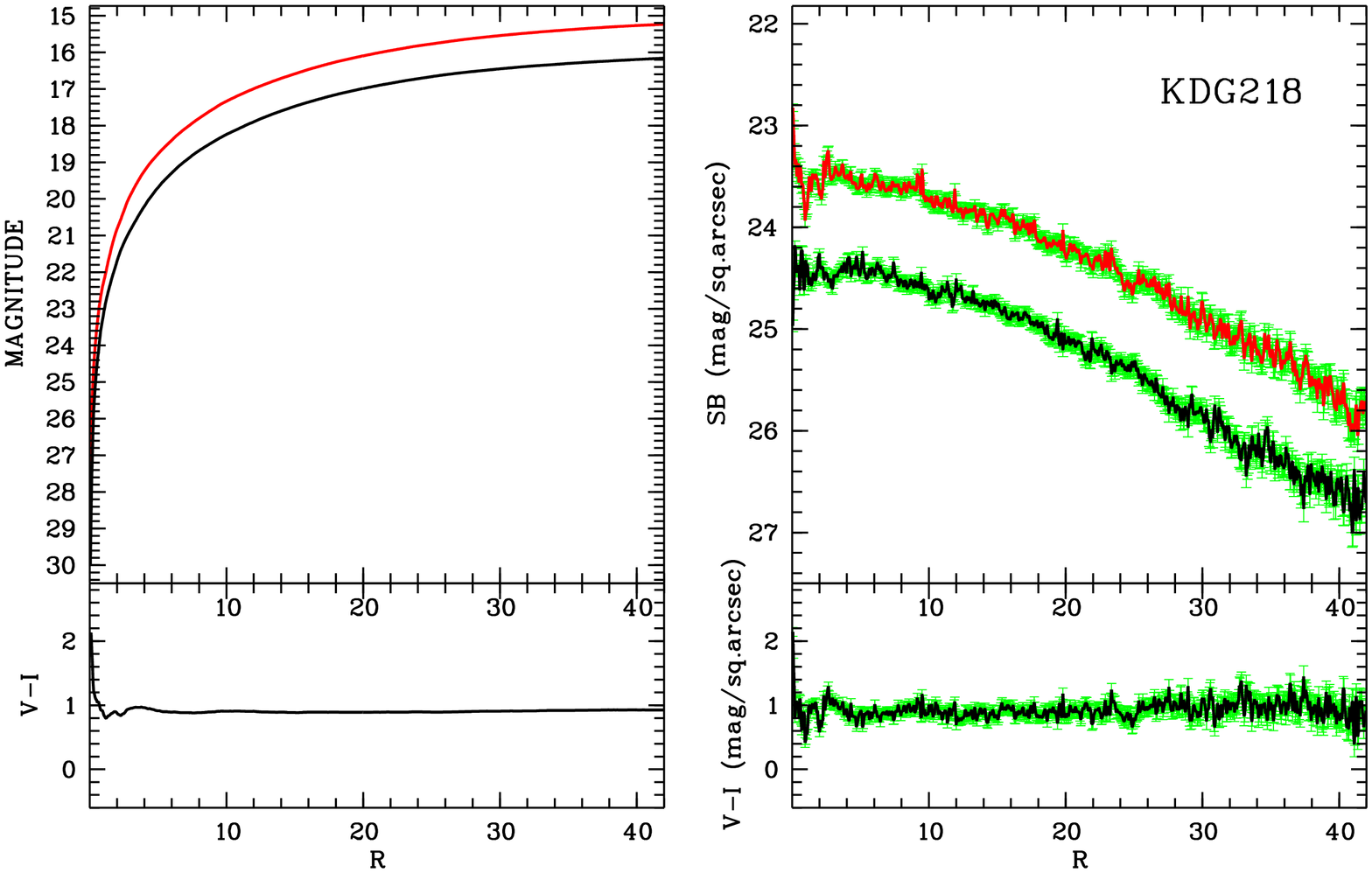}
\caption{Apparent integral magnitude (a) and surface brightness (b)
in the $V$ and $I$ bands in dependence on aperture radius in
arcseconds.} \end{figure*}

We used images of KDG\,218 obtained with the HST/ACS to perform the
surface photometry of the galaxy. Figure~3 shows the results. The
panel a presents behavior of the integral magnitude of KDG\,218 in
the $I$ (1) and $V$ (2) filters depending on the circular radius $R$
(in arcseconds). The bottom panel a reflects the variations with the
radius of the integral color index. From these data, the integral
magnitudes of the galaxy within \mbox{$R=42\arcsec$} are
\mbox{$V(<42\arcsec)=16.18\pm0.05$ mag} and
$I(<42\arcsec)=15.22\pm0.05$~mag. Taking into consideration the
relation $B-V=0.85(V-I)-0.2$~\cite{maka1998:Karachentsev_n_en}, we
obtain \mbox{$B_T\simeq B(<42\arcsec)$}$=16\fm80$ for KDG\,218 that
is in good agreement with the by-eye estimate \mbox{$B_T=16\fm8$}
mentioned above. The galaxy's color index \mbox{$B-V=0.62$}
corresponds to its transition type (Tr) with an older stellar
population. The panel b of Fig.~3 demonstrates the measured profile
of the KDG\,218 brightness in the $I$ and $V$ bands as well as the
variations of the color index $(V-I)$. The central surface brightness
of the galaxy is
$$SB_V(0)=24.39\pm0.05^{\rm m}/\sq\arcsec$$ and
$$SB_I(0)=23.45\pm0.05 ^{\rm m}/\sq\arcsec$$ or  $$SB_B(0)=24\fm99
/\sq\arcsec$$. The brightness profile averaged over two filters
corresponds to the Sersic parameter $n=0.60\pm0.03$ with the
effective radius $R_e=23\farcs6\pm0\farcs6$.

\subsection{Neighborhood and Distance Estimation}
KDG\,218 resides in the region of the so-called ``Virgo Southern
Extension'' at an angular distance of $21\fdg9$ from the center of
the Virgo cluster identified with NGC~4486. This distance corresponds
almost exactly to the radius of the zero-velocity surface $23\fdg6
\pm 2\fdg3$~\cite{kar2014:Karachentsev_n_en} which separates the collapsing region
around the cluster against the global cosmological expansion.
Unfortunately, KDG\,218 still has its radial velocity unmeasured.

There is a spiral late-type galaxy NGC~4948 with the heliocentric
radial velocity $V_h=1123$~km\,s$ ^{-1}$ near KDG\,218 at an angular
separation of 17$\arcmin$. Its distance determined with the SNIa
supernova. In the neighborhood, there are several more galaxies,
distances to which were estimated~\cite{kar+nas2013:Karachentsev_n_en} with the
Tully--Fisher method.

\begin{table*} \setcaptionmargin{0mm} \onelinecaptionstrue \captionstyle{normal}

\caption{Galaxies within $150\arcmin$ of KDG\,218 between $V_h$
$620$ and $1620$~km\,s$ ^{-1}$} \medskip

\begin{tabular}{l|c|c|c|c|c|c|c|c} \hline

\multirow{2}{*}{~~~~~~Name}& \multirow{2}{*}{RA (2000.0) Dec} &SGL,
&SGB, &\multirow{2}{*}{Type} &$V_h$, &$B_T$,& $D$,
&\multirow{2}{*}{Method}\\ &&deg &deg &&km\,s$^{-1}$ &mag &Mpc &\\
\hline

KDG\,218 &$130543.9$ $-074532$ &$124.64$ &$+00.46$ &Tr &~--- &$16.8$
&~--- &\\
DDO\,163 &$130514.3$ $-075321$ &$124.74$ &$+00.31$ &Sm
&$1123$ &$16.0$ &$23.7$ &TF\\
NGC\,4948 &$130455.9$ $-075652$ &$124.77$
&$+00.22$ &Sdm &$1123$ &$14.0$ &$22.0$ &SN\\
2MASX &$130701.6$
$-074155$ &$124.67$ &$+00.79$ &Sdm &$1612$ &$15.1$ &~--- &\\
NGC\,4958 &$130548.9$ $-080113$ &$124.90$ &$+00.41$ &S0 &$1455$
&$12.1$ &$21.4$
&TF\\
$[$KKS\,2000$]$42& $130619.1$ $-080533$ &$125.00$ &$+00.51$ &Tr
&~--- &$18.3$ &~--- &\\
NGC\,4948A &$130505.8$ $-080941$ &$124.99$
&$+00.20$ &Sdm &$1541$ &$14.5$ &$15.6$ &TF\\
NGC\,4951 &$130507.7$
$-062938$ &$123.39$ &$+00.66$ &Scd &$1176$ &$12.6$ &$15.5$ &TF\\
GALEX &$130456.2$ $-094850$ &$126.57$ &$-00.29$ &BCD &$1460$ &$16.3$
&~--- &\\
IC\,4212 &$131203.0$ $-065833$ &$124.32$ &$+02.18$ &Scd
&$1476$ &$14.5$ &$21.9$ &TF\\
IC\,3908 &$125640.6$ $-073346$ &$123.84$
&$-01.65$ &Sd &$1296$ &$13.5$ &$21.6$ &TF\\
NGC\,4818 &$125648.9$
$-083131$ &$124.78$ &$-01.87$ &Sab &$1065$ &$12.1$ &$12.2$ &TF\\
NGC\,4813 &$125636.1$ $-064904$ &$123.12$ &$-01.46$ &Sa &$1394$
&$14.1$
&~--- &\\
RFGC\,2432 &$125848.9$ $-060646$ &$122.59$ &$-00.74$ &Sm
&$1600$ &$14.5$ &$24.2$ &TF\\
NGC\,4941 &$130413.1$ $-053306$ &$122.42$
&$+00.70$ &Sab &$1108$ &$12.4$ &$14.6$ &TF\\
GALEX &$131038.6$
$-095554$ &$127.07$ &$+01.03$ &BCD &$1165$ &$16.5$ &~--- &\\
UGCA\,311 &$125746.8$ $-093801$ &$125.91$ &$-01.94$ &Sdm &$1482$
&$14.2$ &$26.5$ &TF\\ \hline

\end{tabular} \end{table*}

Table~1 presents  equatorial and  supergalactic coordinates as well
as morphological types, heliocentric velocities ($V_h$,~km\,s$
^{-1}$), apparent magnitudes ($B_T$), distances ($D$,~Mpc), and
methods used in their estimation for 16 galaxies with radial
velocities in the range of $V_h=[620, 1620]$~km\,s$ ^{-1}$ within a
radius of 150$\arcmin$ around KDG\,218. Into this list ranged by the
angular distance from KDG\,218, the low-surface-brightness galaxy
[KKS\,2000]42 was included, integral parameters of which are similar
to those of KDG\,218 = [KKS\,2000]41. Figure~4 shows the distribution
of the KDG\,218 neighbors in extragalactic coordinates, where the
sizes of the galaxies are proportional to their apparent magnitude,
and two low-surface-brightness objects are marked with open circles.
The center of the Virgo cluster (SGL$=102\fdg88$, SGB$=-2\fdg35$) is
on the right, far from the figure borders.

\begin{figure} \setcaptionmargin{5mm} \onelinecaptionstrue \captionstyle{normal}
 \includegraphics[scale=0.35,angle=-90]{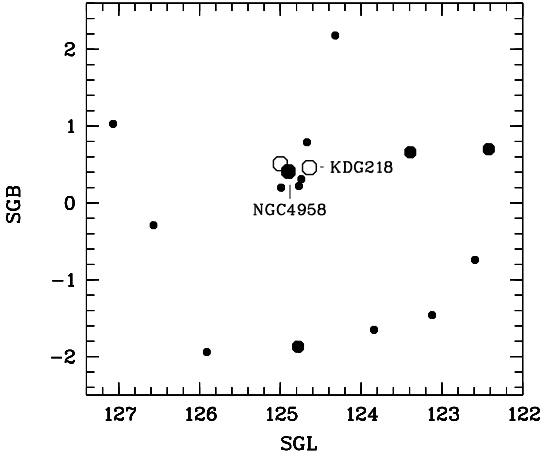}
\caption{Map of the neighborhood of the galaxies with radial
velocities of $V_h=[620-1620]$~km\,s$ ^{-1}$ and  projected
separation less than $2\fdg5$ around KDG\,218.} \end{figure}

According to the given data KDG\,218 along with NGC\,4948 and
DDO\,163 is a possible companion of the massive S0 galaxy NGC\,4958
at a distance of $D\simeq22$~Mpc. In this case, the angular diameter
of KDG\,218 $a_e=0\farcm79$, in which the half of galaxy's luminosity
is included, corresponds to the effective linear diameter
\mbox{$A_e=5.04$} kpc. Consequently, both central surface brightness,
$SB_g(0)=24\fm8/\sq\arcsec$, and effective diameter of KDG\,218
satisfy the criterion of an ultra-diffuse galaxy. This criterion
remains valid if KDG\,218 resides in the VirgoSE filament at the
distance of the Virgo cluster itself ($16.5$~Mpc) and also with the
minimum distance estimate that we obtained $D_{\rm
min(TRGB)}=13.1$~Mpc.

\section{ULTRA-DIFFUSE GALAXIES IN THE LOCAL VOLUME} \subsection{Nearby Ultra-Diffuse Galaxy IKN}

This galaxy is one of the most diffuse in the nearby M\,81 group. It
was resolved into stars with the HST~for the first time
by~\cite{kar2006:Karachentsev_n_en}. The distance measured with the TRGB is $3.75$~Mpc,
confirming the status of IKN as a companion of M\,81. Detailed
surface photometry is not provided for IKN, because a rather bright
star is projected onto its northern side. Okamoto et
al.~\cite{oka2015:Karachentsev_n_en} obtained deep images of the central region of the
M\,81 group at the Hyper Suprime-Cam of the Subaru telescope. The
distribution map of RGB stars (Fig.~5~\cite{oka2015:Karachentsev_n_en}) shows that the
maximum diameter of IKN reaches $7\farcm9$ or $8.6$~kpc. Georgiev et
al.~\cite{geo2009:Karachentsev_n_en} and Tudorica et al.~\cite{tud2015:Karachentsev_n_en} have detected
six globular clusters in IKN, the half of which are concentrated in a
$3.4$~kpc-diameter circle. Taking these data into consideration and
using the photometric profile of IKN on the southern side of the
galaxy from the images obtained with the 6-m telescope~\cite{kar2007:Karachentsev_n_en}
we estimated the effective diameter of IKN equal to $3.15$~kpc. There
is a dwarf galaxy BK5N not far from IKN with the central surface
brightness in $R$ band $SB_R(0)=24\fm5/\sq\arcsec$ according to the
photometry in the MegaCam CFHT images~\cite{chi2009:Karachentsev_n_en}. Having compared
the images of these galaxies, we obtained the central brightness for
IKN by $1\fm5$ fainter which in transition to $g$ band and taking the
Galactic extinction into account yields
\mbox{$SB_g(0)\simeq26\fm9/\sq\arcsec$} for IKN. Consequently, the
effective diameter and central surface brightness of IKN quite well
match its classification as an ultra-diffuse galaxy.

\subsection{Other Nearby Ultra-Diffuse Galaxy Candidates}
\begin{table*} \setcaptionmargin{0mm} \onelinecaptionstrue \captionstyle{normal}

\caption{Ultra-diffuse galaxy candidates in the Local Volume}
\medskip

\begin{tabular}{l|c|l|c|c|c|c|c|c|l|c} \hline

\multirow{2}{*}{~~~~~~~~~Galaxy} &\multirow{2}{*}{RA (2000.0)Dec}
&~~$D$, &$A_e$& \multirow{2}{*}{$b/a$} &$M_B$,& $SB_B(e)$, &$\log
M^*$,& \multirow{2}{*}{$TI_1$} &\multirow{2}{*}{~~~~~MD}
&\multirow{2}{*}{$TI_j$}\\ &&Mpc &kpc &&mag & mag/$\sq\arcsec$
&$M_{\odot}$ &&&\\ \hline

\qquad\quad (1) &(2) &~~(3) &(4) &(5) &(6) &(7) &(8) &(9)& ~~~~~(10)
&(11)\\ \hline

And\,XIX &001932.1+350237 &$0.93$ &$3.52$ &$0.42$ &$-9.6 $ &$29.4$
&$6.80$ &$2.2$ &M\,31 &$1.80$\\ Cas\,III &003559.4+513335 &$0.78$
&$3.60$ &$0.50$ &$-11.5$ &$27.6$ &$7.56$ &$2.3$ &M\,31 &$1.80$\\
Scl-MM-Dw2 &005017.1-244459 &$3.12$ &$5.74$ &$0.34$ &$-11.0$ &$29.1$
&$7.36$ &$0.7$ &NGC\,0253 &$1.76$\\ $[$TT\,2009$]$30 &022254.7+424245
&$9.8 $ &$3.40$ &$0.41$ &$-11.4$ &$27.6$ &$6.79$ &$2.3$ &NGC\,0891
&$1.73$\\ d0226+3325 &022652.8+332537 &$9.5 $ &$4.92$ &$0.92$
&$-12.7$ &$27.1$ &$7.34$ &$3.8$ &NGC\,0925 &$1.07$\\ KK\,69
&085250.7+334752 &$9.16$ &$3.74$ &$0.76$ &$-12.5$ &$26.6$ &$7.27$
&$0.4$ &NGC\,2683 &$1.56$\\ KK\,77 &095010.0+673024 &$3.80$ &$3.15$
&$0.75$ &$-12.2$ &$26.6$ &$7.84$ &$2.5$ &M\,81 &$1.89$\\ GARLAND
&100342.0+684136 &$3.82$ &$4.61$ &$0.60$ &$-11.4$ &$28.2$ &$6.81$
&$3.0$ &NGC\,3077 &$1.89$\\ IKN &100805.9+682357 &$3.75$ &$3.15$
&$0.85$ &$-11.6$ &$27.2$ &$7.60$ &$3.0$ &M\,81 &$1.89$\\ NGC\,3521sat
&110540.7+000715 &$10.7$ &$8.13$ &$0.77$ &$-14.2$ &$26.7$ &$8.62$
&$4.7$ &NGC\,3521 &$1.85$\\ NGC\,4631dw1 &124057.0+324733 &$7.4 $
&$4.71$ &$0.60$ &$-12.6$ &$27.1$ &$7.30$ &$3.0$ &NGC\,4631 &$1.43$\\
CenA-MM-Dw1 &133014.3-415336 &$3.63$ &$3.13$ &$0.81$ &$-12.6$ &$26.2$
&$7.98$ &$2.9$ &NGC\,5128 &$1.88$\\ CenA-MM-Dw3 &133021.5-421133
&$4.61$ &$6.63$ &$0.71$ &$-12.3$ &$28.1$ &$7.88$ &$0.0$ &NGC\,5128
&$1.65$\\ KK\,208 &133635.5-293415 &$5.01$ &$8.77$ &$0.42$ &$-14.4$
&$26.6$ &$8.71$ &$2.7$ &NGC\,5236 &$1.65$\\ Sag\,dSph
&185503.1-302842
&$0.02$ &$5.17$ &$0.48$ &$-12.7$ &$26.7$ &$8.02$ &$5.4$ &MW &$1.80$\\
\hline

Mean &&$5.07$ &$4.82$ &$0.62$ &$-12.2$ &$27.4$ &$7.59$ &$2.6$
&~~~~~~~--- &$1.71$\\ \hline

\end{tabular} \end{table*}

The best way to learn about any galaxy population is to study its
nearest representatives, the structure  and properties of which can
be seen in maximum details. For this purpose, we selected 15 Local
Volume galaxies ($D<11$~Mpc) from the UNGC catalog that match the
determination of an ultra-diffuse galaxy. Unfortunately, the  surface
photometry data for even the nearest galaxies are very few.
Particularly, a number of galaxies have no estimated central surface
brightness and effective diameter. Instead of which we used the
galaxy's Holmberg diameter measured at
\mbox{$SB_B(0)\simeq26\fm5/\sq\arcsec$} and effective surface
brightness. Table~2 shows the summary of parameters of 15
ultra-diffuse galaxy candidates in the Local Volume. Its columns
present: (1, 2)\mbox{---}galaxy's name and its equatorial
coordinates; (3)---distance measured with the TRGB method (2 decimal
places) or from the membership in the group; (4)---effective or
Holmberg diameter; (5)---apparent axial ratio; (6)\mbox{---}absolute
$B$-magnitude corrected for the Galactic extinction; (7)---average
surface brightness in the $B$ band inside the effective radius;
(8)\mbox{---}logarithm of the stellar mass calculated from the $K$
luminosity with \mbox{$M^*/L_K=1.0\times M_{\odot}/L_{\odot}$};
(9)---tidal index from the UNGC determined by the tidal  force of the
nearest and massive neighbor (=Main Disturber), the name of which is
given in column (10); (11)---logarithm of density created by
surrounding galaxies within a $1$~Mpc radius and expressed in units
of average cosmic density. The last row in the table gives average
values of each parameter.

The data from Table~2 allow the first estimation of the abundance of
ultra-diffuse galaxies to be conducted. The total number of the known
galaxies with $D<11$~Mpc is 988, consequently, the relative abundance
of ultra-diffuse galaxies in the Local Volume does not exceed 1.5\%.

Considering the characteristics of the nearest UDG candidates
presented in Table~2, we can conclude on the following.

\begin{list}{}{
\setlength\leftmargin{5mm} \setlength\topsep{1mm}
\setlength\parsep{-0.5mm} \setlength\itemsep{2mm} } \item[1.] The UDG
population is not a homogeneous sample. It contains irregular diffuse
galaxies (Garland, d0226+3325) as well as the objects of the
transition (Tr) type (KK\,69, NGC\,4631dw1, and [TT\,2009]30), in
which star formation is taking place.

\item[2.] Among the diffuse galaxies having an old stellar population
there are elongated objects with the axial ratio $b/a<0.5$
(Sag\,dSph, KK\,208, \mbox{Scl-MM-Dw2}, and And\,XIX). Obviously,
their disturbed shape is caused by tidal action from close massive
neighbor.

\item[3.] The common feature of 15 diffuse galaxies is their location
in high-density regions. According to the UNGC~\cite{kar2013:Karachentsev_n_en}, the
average index $\langle TI_j\rangle=1.71$ means that the local stellar
mass density around them is 50 times higher than the average cosmic
density. The same feature is testified by the high average value of
the index $\langle TI_1\rangle=2.6$ which indicates the presence of a
close massive neighbor near UDG.

\end{list}

The data given in Table~2  leave the question open: if there are any
isolated diffuse galaxies far from massive neighbors, the destroying
tidal influence of which seems the obvious cause of transformation of
a certain dwarf into an extended diffuse stellar flow. To answer this
question, deep surveys of wide sky coverage are needed which are
still absent. In particular, it is interesting to check if there are
any ultra-diffuse galaxies in the local non-virialized cloud of
dwarfs CVnI, the near side of which \mbox{($D\simeq 2.5$~Mpc)} almost
contacts the Local Group.

Among low-surface-brightness galaxies, there is a population of
``nucleated'' ones, where the presence of the ``nucleus'' can
influence the galaxy's classification as an ultra-diffuse one via its
central surface brightness. Alternatively, one can use not the
central but the effective surface brightness $SB(e)$. Then, taking
into consideration the Table~2 data, the criterion of the
ultra-diffuse galaxy takes the form: $$A_e>3.0 \,{\rm kpc},
\,\,\,\,SB_B(e)>26\fm5/\sq\arcsec, \,\,\,\,b/a>0.5.$$ Here we put the
limitation by the apparent axial ratio $b/a$ in order to exclude
trivial cases, when the increase of the effective diameter of a dwarf
galaxy and decrease of its surface brightness are caused by
disruption of its structure with tidal forces.

\section{CONCLUSION}
In recent years, a series of publications appeared reporting on the
existence and properties of a special population of ultra-diffuse
galaxies. Objects of this type with large linear sizes but low
surface brightness are found in clusters and galaxy groups mainly.
According to the papers~\cite{bur2016:Karachentsev_n_en,bur2017:Karachentsev_n_en}, the abundance of UDGs
is the highest in rich clusters. Along with this, they avoid the
densest central region of a cluster and are relatively rare on its
periphery. Thus, a UDG population is an important indicator of
dynamic processes in clusters and groups. From the data
in~\cite{bea2016:Karachentsev_n_en,dok2016:Karachentsev_n_en}, ultra-diffuse galaxies have a high
dark-to-stellar mass ratio, which can reach $M_{\rm DM}/M^*$ about
\mbox{$10^2$--$10^3$} for them.

According to our HST observations, the distance of KDG\,218 is
\mbox{$D>13.1$}~Mpc. It is most likely located in the NGC~4958 group
at a distance of 22~Mpc or in the VirgoSE scattered filament
adjoining the Virgo cluster (16.5~Mpc). In both cases, the effective
diameter of KDG\,218, $A_e>3.0$ kpc, and the central surface
brightness, \mbox{$SB_V(0)=24\fm36/\sq\arcsec$}, correspond to the
criterion of an ultra-diffuse galaxy.

We also give the list of 15 galaxies of the Local Volume that can be
considered as the nearest UDG candidates by their sizes and surface
brightness. The reliable photometry has not been conducted for these
galaxies yet, and they need a detailed study.

\begin{acknowledgments}
The work is supported with the Russian Science Foundation grant
No.~14--12--00965-P.  This work is based on observations made with
the NASA/ESA Hubble Space Telescope, program GO-14636, with data
archive at the Space Telescope Science Institute. STScI is operated
by the Association of Universities for Research in Astronomy, Inc.
under NASA contract NAS 5-26555.
\end{acknowledgments}

{}
\end{document}